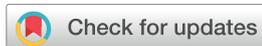

Cite this: DOI: 10.1039/d0sm01192d

# Defective nematogenesis: Gauss curvature in programmable shape-responsive sheets with topological defects†

Daniel Duffy and John S. Biggins *

Flat sheets encoded with patterns of contraction/elongation morph into curved surfaces. If the surfaces bear Gauss curvature, the resulting actuation can be strong and powerful. We deploy the Gauss–Bonnet theorem to deduce the Gauss curvature encoded in a pattern of uniform-magnitude contraction/elongation with spatially varying direction, as is commonly implemented in patterned liquid crystal elastomers. This approach reveals two fundamentally distinct contributions: a structural curvature which depends on the precise form of the pattern, and a topological curvature generated by defects in the contractile direction. These curvatures grow as different functions of the contraction/elongation magnitude, explaining the apparent contradiction between previous calculations for simple +1 defects, and smooth defect-free patterns. We verify these structural and topological contributions by conducting numerical shell calculations on sheets encoded with simple higher-order contractile defects to reveal their activated morphology. Finally we calculate the Gauss curvature generated by patterns with spatially varying magnitude and direction, which leads to additional magnitude gradient contributions to the structural term. We anticipate this form will be useful whenever magnitude and direction are natural variables, including in describing the contraction of a muscle along its patterned fiber direction, or a tissue growing by elongating its cells.

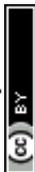

## 1 Introduction

Differential growth and differential muscular contraction underpin many impressive and important shape transformations found in biology,[1] ranging from embryonic gastrulation and limb formation to gut peristalsis and the beat of a heart. The key idea is that, with the right spatial patterning, a limited set of local moves – isotropic growth, directional growth, or uniaxial muscular contraction – can be choreographed into a complex and reliable global shape transformation. Correspondingly, there is considerable interest in developing spatially-programmable shape-changing materials,[2,3] both to elucidate the basic geometric and mechanical principles of differential shape change,[4–6] and for use in soft machines[7,8] and deployable structures.[9]

Several different programmable shape-changing materials have been developed, each responding to different stimuli and offering a different palette of local shape changes. Recent work has highlighted patterned (isotropic) swelling in hydrogels,[10–13] patterned contraction/elongation in liquid crystaline elastomers/glasses (LCE/Gs) subject to heat or light,[14–17] and patterned contraction in "baromorphs" subject to inflation.[18,19] A natural distinction thus arises between a patterned magnitude of a locally isotropic shape change, and a patterned direction of a fixed magnitude shape change (Fig. 1). Returning to biology, although evolution is surely able to pattern magnitudes,[1,20] most biological tissues are anisotropic and also exhibit patterned directions, be they patterns of elongational growth[21,22] or patterns of muscular contraction.

In each of the above responsive materials, shape change is programmed into an initially flat sheet, which, on activation, morphs into a curved surface. Strikingly, such transformations are often impossible with passive sheets, generating timeless struggles for tailors, mapmakers, architects, gift-wrappers and manufacturers. As immortalized in Gauss's Theorema Egregium,[23,24] the key difficulty is geometric: the Gauss curvature of a surface (calculated as the product of the two principal curvatures, $K = 1/(R_1 R_2)$) cannot be modified without changing the in-surface distance between points, encoded by the surface's metric. Thus a flat sheet of paper can be bent into a cylinder, but cannot be coerced into a sphere. Active shape-changing sheets, like their biological counterparts, can side-step this geometric constraint as their programmed shape-change does indeed change the in-surface distance between points.[10,25,26] Directional growth offers a unique additional possibility: patterns with topological defects. Such directional defects are familiar from liquid nematics[27] and have inspired many patterns encoded in LCE/Gs.[26,28,29] Indeed, LCE sheets encoded with concentric circles of contraction, forming a +1 nematic defect,

Engineering Dept., University of Cambridge, Trumpington St., Cambridge, CB2 1PZ, UK. E-mail: jsb56@cam.ac.uk
† Electronic supplementary information (ESI) available. See DOI: 10.1039/d0sm01192d

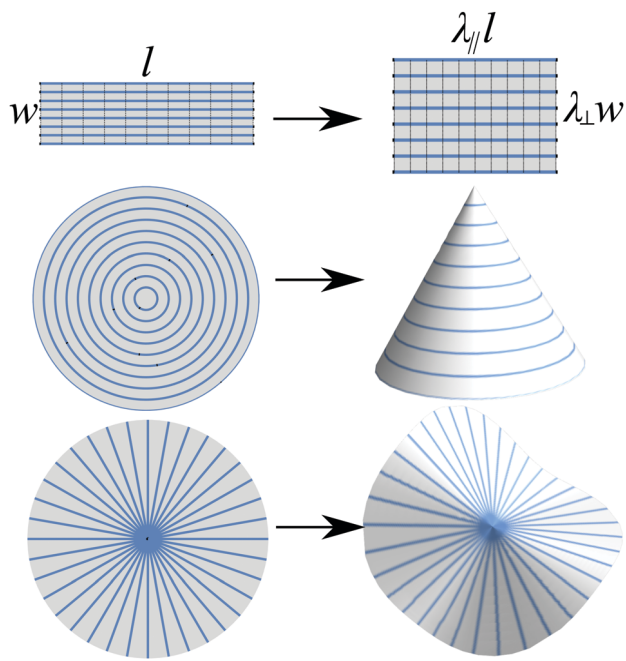

Fig. 1 A sheet of shape-shifting material contracts along a programmed direction (blue lines) on activation. If the direction is constant the sheet remains flat (top). Concentric rings of contraction morph a disk into a cone, while a radial contraction pattern makes an anticone.

have become a ubiquitous test-case for new patterning techniques.[15,28,30–33] Such sheets morph dramatically into cones (Fig. 1) demonstrating that topological defects offer a way to produce sharp features with divergent Gauss curvature.[26]

Conversely, Gauss's geometric coupling of bend and stretch make curved surfaces strong and mechanically favourable. For example, if a flat sheet is bent in one direction, it cannot then bend in the transverse direction without stretch.[34] This curvature-induced rigidity, which is familiar from corrugated cardboard and pizza consumption,[35] leads to the general result that, whilst flat sheets buckle in a pure bending mode at a characteristic compressive force $F \propto t^3$, curved shells require both stretch and bend to buckle, leading to a much higher buckling force $F \propto t^2$ dictated by the harmonic mean of the bending and stretching moduli.[36–38] Consequently, when an active sheet morphs into a Gauss-curved surface, the transition is mechanically strong: the LCE defect cones will lift 2500 times their own weight[32] as they activate. Patterns of Gauss curvature are also an essential component of inverse design: in order to program a sheet to morph into a desired shape it is necessary (though not sufficient) that the programmed shape change produces the desired Gauss curvature.[11,16,28,39–41]

In this paper, we use the Gauss–Bonnet theorem to calculate the distribution of Gauss curvature in an initially flat sheet programmed with a directional pattern of shape change. Our calculation is greatly facilitated by the use of the natural coordinate system of a nematic field, recently introduced by Niv and Efrati.[42] The Gauss–Bonnet approach allows us to construct a single result that unifies previous work on distributed Gauss curvature[39,40,43,44] and sharp points with divergent Gauss curvature.[26,45] Our unified result reveals a delicate interplay between topology and spatial patterning in the resultant Gauss curvature, and allows us to calculate the Gauss curvature encoded in topological defects with charges other than +1. We extend and verify these results by conducting numerical shell calculations on sheets encoded with higher-order defects to clarify the full 3D form of the surface that emerges. Finally, we present an analogous result for sheets encoded with directional growth in which both the magnitude and direction vary spatially in plane. Although such patterning is yet to be demonstrated in liquid crystalline solids, we anticipate that it will soon (either via local crosslink density or via control of the imprinted order parameter), and has already been achieved by evolution in many different biological contexts.

## 2 Curvature induced in flat sheets with a programmed direction of contraction/elongation

We first consider a planar sheet, in which each point is programmed with a planar direction $\boldsymbol{n} = (\cos\psi, \sin\psi)$ along which the material will contract/elongate by a factor of $\lambda_\parallel$ upon activation, accompanied by, a sympathetic change of $\lambda_\perp$ in the perpendicular direction. In an LCE/G, $\boldsymbol{n}$ would correspond to the nematic director (alignment direction), and, as indicated in Fig. 1, the response on heating/illumination is a large contraction, $0.25 < \lambda_\parallel < 1$, while the lateral expansion, $\lambda_\perp = \lambda_\parallel^{-\nu}$, is determined by the opto-thermal Poisson ratio, which is strictly $\nu = 1/2$ in incompressible elastomers, but can be as high as 2 in photo-glasses. In baromorphs, $\boldsymbol{n}$ is the programmed pneumatic channel direction, and on inflation the channel does not change in length, $\lambda_\parallel = 1$, but ideally contracts laterally by $\lambda_\perp = 2/\pi = 0.63\ldots$.

As shown in Fig. 2, we may also define an orthogonal dual director on the sheet with $\psi \to \psi + \pi/2$, such that $\boldsymbol{n}^\star = (-\sin\psi, \cos\psi)$. Then, in the undistorted sheet, the infinitesimal length element $\mathrm{d}\boldsymbol{l} = (\mathrm{d}x, \mathrm{d}y)$ has length $\mathrm{d}l^2 = \mathrm{d}\boldsymbol{l}^\mathrm{T} I \mathrm{d}\boldsymbol{l}$, which, on activation, becomes

$$\mathrm{d}l_\mathrm{A}^2 = \mathrm{d}\boldsymbol{l}^\mathrm{T}(\lambda_\parallel^2 \boldsymbol{nn} + \lambda_\perp^2 \boldsymbol{n}^\star\boldsymbol{n}^\star)\mathrm{d}\boldsymbol{l} \equiv \mathrm{d}\boldsymbol{l}^\mathrm{T} \bar{a} \mathrm{d}\boldsymbol{l}.$$

The activated sheet (indicated via subscript A) must deform into a surface following the programmed metric, $\bar{a}$. Suitable spatial variation in $n$ will generate a spatially varying metric bearing Gauss curvature, guaranteeing that the resultant surface will also be curved.

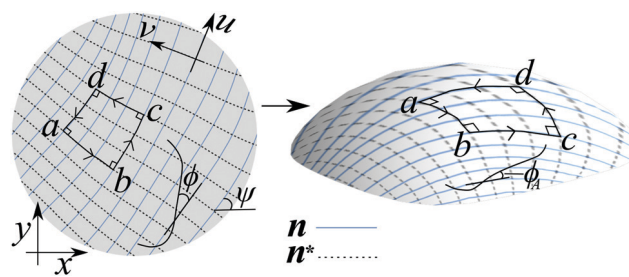

Fig. 2 An initially flat patch of nematic elastomer encoded with a spatially varying director $\boldsymbol{n}$ (left) will, upon activation, contract everywhere along $\boldsymbol{n}$ and expand along $\boldsymbol{n}^\star$, causing it to morph into a curved surface (right). The nematic pattern defines a natural orthogonal coordinate system for the patch, $(u,v)$, with $u$-lines along $\boldsymbol{n}$ and $v$-lines along $\boldsymbol{n}^\star$.

Our task is to compute the intrinsic geometry of the resultant surface, as characterized by its geodesic and Gauss curvatures. The metric is invariant under $n \to -n$, which is an inevitable consequence of the quadrupolar nature of nematic order. However, even if the underlying order really were vectorial, the metric, as a quadratic form, would nevertheless have this quadrupolar symmetry. Thus nematic rather than vectorial patterns are the natural language of directional shape change, and nematic defects, including half integer defects, could be found in any directional shape-changing system.

Previous authors have applied the Theorema Egregium to directly calculate Gauss curvature from gradients of the metric, yielding[40,43,44]

$$K_A = \frac{1}{2}(\lambda_\perp^{-2} - \lambda_\parallel^{-2}) \left[ (\partial_y^2 \psi - \partial_x^2 \psi - 4\partial_x \psi \partial_y \psi) \sin 2\psi \right. \\ \left. + 2(\partial_x \partial_y \psi + (\partial_y \psi)^2 - (\partial_x \psi)^2) \cos 2\psi \right]. \quad (1)$$

This form has been used successfully to design surfaces with uniform finite Gauss curvature (such as spherical caps) and arbitary surfaces of revolution,[46,47] and underpins recent work on inverse design.[40]

However, if we attempt to apply the above derivative formula at a topological defect in the pattern we encounter a problem: the derivatives are divergent. To clarify this, let us first consider two well-understood +1 defect nematic director patterns: those with the director in concentric circles around a single point, and those with the director emanating radially from a single point, each shown in Fig. 1. If a disk is encoded with concentric circles and then activated, it forms a new surface where a circle with original radius $r$ has new (in material) radius $r\lambda_\perp > r$, and new circumference $2\pi r\lambda_\parallel < 2\pi r$. This geometric contradiction is resolved by the disk morphing into a cone, with half-angle $\sin^{-1}(\lambda_\parallel/\lambda_\perp)$. The sides of the conical surface are Gauss flat everywhere, $K_A = 0$, in agreement with a direct application of eqn (1). However, at the tip of the cone the Gauss curvature is clearly infinite. Furthermore, this tip of infinite Gauss curvature makes a finite contribution to the integrated (aka total) curvature, $\int K_A dA_A = 2\pi(1 - \lambda_\parallel/\lambda_\perp) > 0$, as may be computed by regularizing the tip with a spherical-cap: the cone's Gauss curvature is a delta function at the tip, with an infinite value but a finite integral. In contrast, the radial defect pattern creates a surface where the circumferences are too long for the radii, which buckles out of plane into an "anticone" shape resembling a saddle. This surface is also Gauss flat, except for an infinite negative curvature at the center which, since this is equivalent to exchanging $\lambda_\parallel$ and $\lambda_\perp$, gives a finite integrated curvature $\int K_A dA_A = 2\pi(1 - \lambda_\perp/\lambda_\parallel) < 0$.[26,48]

With suitable care, differentiation can yield delta functions; for example in 3D electrostatics we routinely write $\nabla^2 r^{-1} = -4\pi\delta^3(r)$. Thus one might reasonably hope to derive these cone/anticone tip curvatures via a careful application of eqn (1). However, any such attempt is doomed to fail. Eqn (1) has the general property that the Gauss curvature encoded by an orthogonal dual pattern ($\psi \to \psi + \pi/2$) is the negative of that of the original pattern, $K_A \to -K_A$. In contrast, although the cone and anticone are orthogonal duals, and do have opposite signs of integrated Gauss curvature, the magnitudes are not equal.

A new approach is provided by the Gauss–Bonnet formula,[24] which relates the integrated Gauss curvature over a surface $S$, the geodesic curvature ($k_g$) of the surface's boundary, and the topological classification of the region via its Euler Characteristic $\chi(S)$:

$$\int_S K dA + \int_{\partial S} k_g ds = 2\pi \chi(S).$$

Geodesic curvature, like Gauss curvature, is an intrinsic property of a path on a surface, and can be computed directly from the metric. We may use Gauss–Bonnet to calculate $\int K dA$ for any patch of surface from the geodesic curvature of the patch boundary. Gauss–Bonnet thus provides a rigorous basis for the assignation of finite integrated curvature to the tip of an (anti)cone, since it can be inferred from the finite geodesic curvature of an encircling path. Furthermore, we may slit a cone along a straight generator and unroll it isometrically into the plane to form a sector of a disk. Applying Gauss–Bonnet in the unrolled state, we then see that the tip's finite integrated curvature is exactly equal to the angular deficit of the sector, which becomes an angular surplus in the anticone case.

In general, any such point with finite integrated curvature is associated with a sharp (aka singular) point in the surface with a discontinuous surface normal. Unlike many other sharp points, such as those created in paper-folding origami, these sharp points cannot be removed by isometric bending deformations, as integrated curvature is a property of the metric alone. We thus describe them as intrinsically sharp points of the surface, and interpret the finite integrated curvature as an angular deficit/surplus associated with the point itself. Away from such points, where $K$ is finite, we may apply Gauss–Bonnet to an infinitesimal patch (over which $K$ is effectively constant) to infer the value of $K$ itself. We will term the finite $K$ in such regions distributed Gauss curvature.

Returning to nematogenesis, our general approach is thus to consider a region of patterned nematic sheet, bounded by a curve along which the nematic director is smooth, compute the geodesic curvature along this path, and then use Gauss–Bonnet to deduce the integrated Gauss curvature within. This approach will work even if the region contains topological defects, and whether or not the activated region contains intrinsically sharp points. Since we may choose any patch of surface, we will thus be able to identify the value of $K_A$ in regions with distributed finite curvature, and the finite integrated curvature $\int K_A dA_A$ at intrinsically sharp points where $K_A$ itself is infinite.

### 2.1 Geodesic curvature

The first step is to compute the geodesic curvature in the activated surface, of a path $\mathbf{r}(l) = (x(l), y(l))$, defined in the flat sheet. In general this is a very involved calculation, but given an orthogonal coordinate system $(u,v)$, in which the metric will take the simplified form diag $(E,G)$, the geodesic curvature for the unit-speed curve $(u(l),v(l))$ takes the simple form ([24] p. 296) sometimes known as Liouville's formula:

$$k_g = \phi' - \frac{1}{2\sqrt{EG}}(u'E_v - v'G_u). \quad (2)$$

Above, $\phi$ is the angle between the curve tangent and the $u$-line tangent in the tangent plane of the surface (as shown in Fig. 2).

Following Niv and Efrati,[42] we thus move into the nematic pattern's natural coordinate system $(u,v)$ in which $u$-lines (i.e., $v$ = const.) are tangent to $\boldsymbol{n}$ and $v$ lines ($u$ = const.) are tangent to $\boldsymbol{n}^*$ (Fig. 2). Since this parameterization is locally orthogonal, the metric of the flat (non-activated) sheet will be diagonal,

$$dl^2 = (du, dv)\begin{pmatrix} \eta^2 & 0 \\ 0 & \beta^2 \end{pmatrix}\begin{pmatrix} du \\ dv \end{pmatrix}, \quad (3)$$

where the scalar fields $\eta(u,v)$ and $\beta(u,v)$ are computed from the geometry of the pattern in question.[42] Furthermore, given the principal stretches on activation are also along $\boldsymbol{n}$ and $\boldsymbol{n}^*$, the $u$ and $v$ lines will also be orthogonal on the deformed surface, and the (diagonal) activated metric is simply[40]

$$dl_A^2 = (du, dv)\begin{pmatrix} \lambda_\parallel^2 \eta^2 & 0 \\ 0 & \lambda_\perp^2 \beta^2 \end{pmatrix}\begin{pmatrix} du \\ dv \end{pmatrix}.$$

These activated $u$ and $v$ lines thus also describe a nematic field (and its orthogonal dual) on the deformed surface. Given the activating deformation is continuous, this nematic field has the same topology as the reference state field, including any topological defects. A direct application of eqn (2) reveals that the integrated geodesic curvature along a path $(u(l),v(l))$ in the activated surface is

$$\int k_{gA} dl_A = \int d\phi_A + \int \left(\frac{\lambda_\perp}{\lambda_\parallel}\frac{\beta_u}{\eta\beta}\beta dv - \frac{\lambda_\parallel}{\lambda_\perp}\frac{\eta_v}{\eta\beta}\eta du\right). \quad (4)$$

Examining Fig. 2, the angle $\phi$ between the path and a $u$-line in the unactivated flat (aka reference) state is given by $\tan\phi = \frac{\delta v}{\delta u}$ which, upon activation becomes $\tan\phi_A = \frac{\lambda_\perp \delta v}{\lambda_\parallel \delta u} = (\lambda_\perp/\lambda_\parallel)\tan\phi$. This allows us to compute the $\phi_A$ term in the geodesic curvature as a reference-state integral

$$\int d\phi_A = \int d(\tan^{-1}[(\lambda_\perp/\lambda_\parallel)\tan\phi])$$
$$= \int \frac{\lambda_\parallel \lambda_\perp}{\lambda_\parallel^2 \cos^2\phi + \lambda_\perp^2 \sin^2\phi} d\phi.$$

The second term of eqn (4) can also be evaluated in the reference state by recognizing that $d\boldsymbol{l} = \eta\, du\,\boldsymbol{n} + \beta\, dv\,\boldsymbol{n}^*$:

$$\int k_{gA} dl_A = \int d\phi_A + \int\left(\frac{\lambda_\perp}{\lambda_\parallel}\frac{\beta_u}{\eta\beta}\boldsymbol{n}^* - \frac{\lambda_\parallel}{\lambda_\perp}\frac{\eta_v}{\eta\beta}\boldsymbol{n}\right)\cdot d\boldsymbol{l}. \quad (5)$$

Along a $u$ line or a $v$ line we have $\phi_A' = 0$, as the angle between the curve tangent and the coordinate line is fixed. Setting $\lambda_\parallel = \lambda_\perp = 1$ to interrogate the non-activated flat sheet, we see that the curvature of a $u$-line is $-\eta_v/(\eta\beta)$, and the curvature of the $v$-line is

$\beta_u/(\eta\beta)$. However, these curvatures are simply the 2D bend and splay (i.e., 2D curl and divergence) of the flat nematic pattern,[42]

$$b = \nabla \times \boldsymbol{n} = \boldsymbol{n}\cdot\nabla\psi = -\frac{\eta_v}{\eta\beta}, \quad (6)$$

$$s = \nabla \cdot \boldsymbol{n} = \boldsymbol{n}^*\cdot\nabla\psi = \frac{\beta_u}{\eta\beta}, \quad (7)$$

which allows us to eliminate the unknown fields $\eta$ and $\beta$ from eqn (5), to get a coordinate-free result for the activated geodesic curvature along any path:

$$\int k_{gA} dl_A = \int \frac{\lambda_\parallel \lambda_\perp}{\lambda_\parallel^2 \cos^2\phi + \lambda_\perp^2 \sin^2\phi} d\phi + \int\left(\frac{\lambda_\perp}{\lambda_\parallel} s\boldsymbol{n}^* + \frac{\lambda_\parallel}{\lambda_\perp} b\boldsymbol{n}\right)\cdot d\boldsymbol{l}. \quad (8)$$

This integrated geodesic curvature is appropriate for Gauss–Bonnet. To evaluate the local value, $k_{gA}$ one may equate integrands of eqn (8), whilst accounting for the infinitesimal arc length, $dl_A = \sqrt{\lambda_\parallel^2 \cos^2\phi + \lambda_\perp^2 \sin^2\phi}\, dl$ to get

$$k_{gA} = \frac{\lambda_\parallel \lambda_\perp}{(\lambda_\parallel^2 \cos^2\phi + \lambda_\perp^2 \sin^2\phi)^{3/2}}\frac{d\phi}{dl}$$
$$+ \frac{(\lambda_\perp/\lambda_\parallel)s\sin\phi + (\lambda_\parallel/\lambda_\perp)b\cos\phi}{\sqrt{\lambda_\parallel^2 \cos^2\phi + \lambda_\perp^2 \sin^2\phi}}. \quad (9)$$

### 2.2 Gauss curvature away from topological defects

We now apply Gauss–Bonnet to a closed loop built from alternating $u$ and $v$ line segments. Along each segment $d\phi_A = d\phi = 0$, so the only contribution to $k_{gA}$ comes from the final term in eqn (8), whilst at each corner there is a concentrated contribution $\int d\phi_A = \pm\pi/2$, since the $u$ and $v$ lines meet perpendicularly even in the activated surface. In the simplest case, we combine two $u$ lines and two $v$ lines to form a quadrangular region, as shown in Fig. 2, giving four $+\pi/2$ contributions in the Gauss–Bonnet formula, which, given the region is topologically a disk with $\chi = 1$, exactly cancel the $2\pi\chi$.

$$\int K_A dA_A = -\oint\left(\frac{\lambda_\perp}{\lambda_\parallel}s\boldsymbol{n}^* + \frac{\lambda_\parallel}{\lambda_\perp}b\boldsymbol{n}\right)\cdot d\boldsymbol{l}. \quad (10)$$

The value of this integral is unchanged by an equal rotation of each vector. Taking a clockwise $\pi/2$ rotation, the line element rotates to point along the loop's outward normal, $d\boldsymbol{l} \to \hat{\boldsymbol{v}}dl$, so the integral is re-cast as a boundary flux. This rotation also transforms $\boldsymbol{n} \to -\boldsymbol{n}^*$ and $\boldsymbol{n}^* \to \boldsymbol{n}$, yielding:

$$\int K_A dA_A = -\oint\left(\frac{\lambda_\perp}{\lambda_\parallel}s\boldsymbol{n} - \frac{\lambda_\parallel}{\lambda_\perp}b\boldsymbol{n}^*\right)\cdot \hat{\boldsymbol{v}}dl.$$

We further simplify this expression by introducing the bend and splay vectors, $\boldsymbol{b} = b\boldsymbol{n}^*$ and $\boldsymbol{s} = s\boldsymbol{n}$, respectively the curvature vector of $u$ lines and minus the curvature vector of $v$ lines. Unlike the director fields themselves, these are true nematic objects (invariant under $\mathbf{n} \to -\mathbf{n}$) and lead to a particularly natural expression:

$$\int K_A dA_A = \oint \left(\frac{\lambda_\parallel}{\lambda_\perp}\mathbf{b} - \frac{\lambda_\perp}{\lambda_\parallel}\mathbf{s}\right) \cdot \hat{\mathbf{v}} dl. \quad (11)$$

In a region without intrinsically sharp points, we may now apply the divergence theorem to identify the local value of the Gauss curvature

$$\int K_A dA_A = \int \nabla \cdot \left(\frac{\lambda_\parallel}{\lambda_\perp}\mathbf{b} - \frac{\lambda_\perp}{\lambda_\parallel}\mathbf{s}\right) dA.$$

Although this was derived for a quadrangular region bounded by $u$ and $v$ lines, it applies to any shape of region, as can be seen by tiling with infinitesimal quadrangles. Expanding the divergence, and recognizing that $dA_A = \lambda_\parallel \lambda_\perp dA$, we find that the local value of Gauss curvature is $K_A = \lambda_\perp^{-2}(-b^2 + (\mathbf{n}^\star \cdot \nabla)b) + \lambda_\parallel^{-2}(-s^2 - (\mathbf{n}\cdot\nabla)s)$, in exact agreement with[42] and.[40] Given the dependence on and $\lambda_\parallel$ and $\lambda_\perp$, this form doesn't immediately resemble eqn (1), although direct substitution of (smooth) $b$, $s$, $\mathbf{n}$ and $\mathbf{n}^\star$ reveals the two do indeed agree.

### 2.3 Gauss curvature with topological defects

However, the above analysis fails if there is a topological defect within the region because, as seen in Fig. 3, the simplest region is no longer quadrangular: it will require different numbers of sides depending on the topological charge, $m$, enclosed. Working in the reference domain, the tangent vector must wind by $2\pi$ around the closed loop. Of these, it will wind by $2\pi m$ in the $u$ and $v$ line segments, and hence the contribution from the corners must be $2\pi(1 - m)$. This introduces an additional $2\pi m$ into eqn (11) leading to

$$\int K_A dA_A = \oint \left(\frac{\lambda_\parallel}{\lambda_\perp}\mathbf{b} - \frac{\lambda_\perp}{\lambda_\parallel}\mathbf{s}\right) \cdot \hat{\mathbf{v}} dl + 2\pi \sum_i m_i,$$

where the $m_i$ are the topological charges enclosed. This form can be simplified by noting that $\nabla\psi = b\mathbf{n} + s\mathbf{n}^\star$, and hence

$$2\pi m = \oint (b\mathbf{n} + s\mathbf{n}^\star) \cdot d\mathbf{l} = -\oint (\mathbf{b} - \mathbf{s}) \cdot \hat{\mathbf{v}} dl, \quad (12)$$

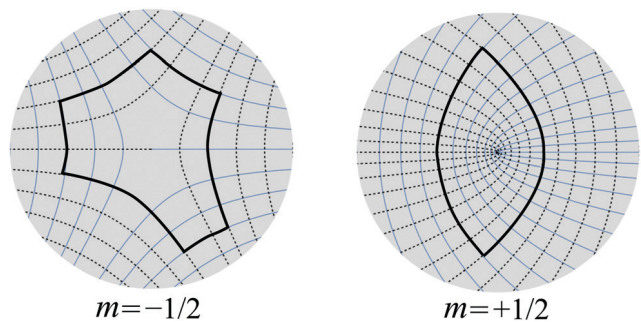

Fig. 3 Examples of defect patterns with topological charge $\pm 1/2$. Note that a path around the defect consisting of $u$ and $v$ segments must have different numbers of segments for each charge.

so we can integrate the $\mathbf{b} - \mathbf{s}$ component of our previous result, to get

$$\int K_A dA_A = \frac{1}{2}\left(\frac{\lambda_\parallel}{\lambda_\perp} - \frac{\lambda_\perp}{\lambda_\parallel}\right) \oint (\mathbf{b} + \mathbf{s}) \cdot \hat{\mathbf{v}} dl$$
$$+ \sum_i m_i \pi \left(1 - \frac{\lambda_\perp}{\lambda_\parallel}\right)\left(1 - \frac{\lambda_\parallel}{\lambda_\perp}\right). \quad (13)$$

This formula is our main result for the Gauss curvature. We will call the first term the structural curvature, since it depends on the local structure of the pattern, while the second term is naturally called the topological curvature. The topological curvature contributes to any patch containing the defect, so it is a finite contribution to the integrated curvature at the point of the defect. Importantly, this does not imply that all defects with the same charge encode intrinsically sharp points with the same integrated curvature, as the structural term may also make a contribution potentially canceling or even reversing the topological one. If we are away from any defect, we can again apply the divergence theorem and equate integrands to compute the local distributed Gauss curvature. Recognizing that $dA_A = \lambda_\parallel \lambda_\perp dA$, this gives

$$K_A = \frac{1}{2}(\lambda_\perp^{-2} - \lambda_\parallel^{-2})\nabla \cdot (\mathbf{b} + \mathbf{s}),$$

which manifestly has the same dependence on $\lambda_\parallel$ and $\lambda_\perp$ as eqn (1), and matches exactly upon substitution for $\mathbf{b}$ and $\mathbf{s}$.

Alternatively, one may use the gradient result to bring all the topological terms into the integral, which leads to a particularly compact form for the Gauss curvature:

$$\int K_A dA_A = \oint \left(\left(\frac{\lambda_\parallel}{\lambda_\perp} - 1\right)\mathbf{b} + \left(1 - \frac{\lambda_\perp}{\lambda_\parallel}\right)\mathbf{s}\right) \cdot \hat{\mathbf{v}} dl. \quad (14)$$

In both cases, the new forms are equivalent to the original result for distributed Gauss curvature in the absence of defects, but also give the correct result for a region containing defects provided the director is smooth on the boundary.

## 3 Constant-rotation $m = 1$ defects

We first test our new form on the familiar constant-rotation +1 defect patterns[26] which, in plane polar coordinates $(r,\theta)$ have the form $\psi = \theta + \alpha$ and hence $\nabla\psi = \mathbf{e}_\theta/r$, $\mathbf{n} = \cos(\alpha)\mathbf{e}_r + \sin(\alpha)\mathbf{e}_\theta$, and $\mathbf{n}^\star = -\sin(\alpha)\mathbf{e}_r + \cos(\alpha)\mathbf{e}_\theta$. The director curves thus form log-spirals with constant angle $\alpha$ between the director and the radial direction, such that $\alpha = 0$ is a radial defect and $\alpha = \pi/2$ produces circles. In this case the bend and splay are given by

$$\mathbf{b} = \frac{\sin\alpha}{r}\mathbf{n}^\star, \quad \mathbf{s} = \frac{\cos\alpha}{r}\mathbf{n}. \quad (15)$$

In the particular case of a circular pattern, the splay vanishes and $\mathbf{b} = -\mathbf{e}_r/r$, while a radial pattern has vanishing bend and $\mathbf{s} = \mathbf{e}_r/r$. Applying eqn (14) to a circular patch centered on the origin, we can calculate the integrated Gauss curvature as

$$\int K_A dA_A = -2\pi\left(\frac{\lambda_\parallel}{\lambda_\perp} - 1\right)\sin^2(\alpha) + 2\pi\left(1 - \frac{\lambda_\perp}{\lambda_\parallel}\right)\cos^2(\alpha)$$
$$= 2\pi\left(1 - \sin^2(\alpha)\frac{\lambda_\parallel}{\lambda_\perp} - \cos^2(\alpha)\frac{\lambda_\perp}{\lambda_\parallel}\right).$$

Since this is independent of the radius of the patch (and the patterns have rotational symmetry), we conclude that there is no distributed Gauss curvature, but there is an intrinsically sharp point at the origin with finite integrated curvature. Equivalently, we can reach this conclusion directly by applying the divergence theorem to eqn (14) and recalling that $\nabla \cdot (\mathbf{e}_r/r) = 2\pi\delta^2(\mathbf{r})$ and $\nabla \cdot (\mathbf{e}_\theta/r) = 0$, so $\nabla \cdot \mathbf{b} = -2\pi\delta^2(\mathbf{r})\sin^2\alpha$ and $\nabla \cdot \mathbf{s} = 2\pi\delta^2(\mathbf{r})\cos^2\alpha$.

We can compute the same result from eqn (13) by noting that $\mathbf{b} + \mathbf{s} = \frac{1}{r}(\cos(2\alpha)\mathbf{e}_r + \sin(2\alpha)\mathbf{e}_\theta)$ to get:

$$\int K_A dA_A = \pi\left(\frac{\lambda_\parallel}{\lambda_\perp} - \frac{\lambda_\perp}{\lambda_\parallel}\right)\cos(2\alpha) + \pi\left(1 - \frac{\lambda_\perp}{\lambda_\parallel}\right)\left(1 - \frac{\lambda_\parallel}{\lambda_\perp}\right)$$
$$= 2\pi\left(1 - \sin^2(\alpha)\frac{\lambda_\parallel}{\lambda_\perp} - \cos^2(\alpha)\frac{\lambda_\perp}{\lambda_\parallel}\right).$$
(16)

Either way, setting $\alpha = 0, \pi/2$ we recapitulate the familiar results for anticones/cones formed from radial/circumferential patterns respectively. However, for an intermediate angle, $\cos(2\alpha_c) = (\lambda_\parallel - \lambda_\perp)/(\lambda_\parallel + \lambda_\perp)$, the integrated curvature at the origin is zero, so there is no intrinsically sharp point in the activated surface; in fact, given the distributed curvature is also zero everywhere in the flanks, the activated nematic sheet will be completely flat.[26]

The second approach clarifies that the integrated curvature at the origin contains a topological contribution, which is identical for all $m = 1$ defects, and a pattern-dependent structural contribution $\propto \cos(2\alpha)$ which is not. The structural contribution reverses sign under taking the orthogonal dual, while the topological contribution does not, leading circular and radial patterns to have different but not exactly opposite curvature. Perhaps surprisingly, the structural term contributes a finite integrated curvature at the origin, despite the distributed curvature being zero throughout the flanks.

## 4 Integrated curvature at a general nematic defect

We next consider the neighborhood of a general defect with topological charge $m$. Constructing a polar coordinate system centered on the defect, the director will have the form $\psi = \theta + \alpha(\theta,r)$, where $\alpha$ is now a varying function that will wind $m - 1$ times on a path around the origin. Computing the bend and splay, we now get

$$b = \frac{\sin\alpha}{r}\left(1 + \frac{\partial\alpha}{\partial\theta}\right) + \frac{\partial\alpha}{\partial r}\cos\alpha \quad (17)$$

$$s = \frac{\cos\alpha}{r}\left(1 + \frac{\partial\alpha}{\partial\theta}\right) - \frac{\partial\alpha}{\partial r}\sin\alpha, \quad (18)$$

so, defining $\alpha_0(\theta) = \alpha(\theta,0)$, near the defect we have

$$\mathbf{b} + \mathbf{s} = \frac{1 + \alpha_0'}{r}(\cos(2\alpha_0)\mathbf{e}_r + \sin(2\alpha_0)\mathbf{e}_\theta) + O(1). \quad (19)$$

Considering a small circular patch at radius $r$, we can again use eqn (13), to find the finite integrated curvature at the defect point,

$$\int K_A dA_A = \pi\left(\frac{\lambda_\parallel}{\lambda_\perp} - \frac{\lambda_\perp}{\lambda_\parallel}\right)\langle\cos(2\alpha_0)\rangle + m\pi\left(1 - \frac{\lambda_\perp}{\lambda_\parallel}\right)\left(1 - \frac{\lambda_\parallel}{\lambda_\perp}\right), \quad (20)$$

where the angle brackets indicate an angular average around the defect. This result is strongly reminiscent of eqn (16), and again is the sum of a structural part, which depends on the particular pattern of the defect, and a topological part which depends only on its charge. Given that the factor $\left(1 - \frac{\lambda_\perp}{\lambda_\parallel}\right)\left(1 - \frac{\lambda_\parallel}{\lambda_\perp}\right)$ is strictly negative, we conclude that positive charge defects are inclined to produce negative Gauss curvature, and vice versa. Furthermore, given $-1 < \langle\cos(2\alpha_0)\rangle < 1$, the structural contribution provides an identical range of curvatures in all defect charges, while the topological term produces an offset to this range proportional to the defect charge. In particular, a higher-order defect will have zero integrated curvature, and hence will not generate an intrinsically sharp point in the surface, if

$$\langle\cos(2\alpha_0)\rangle = m\frac{\lambda_\parallel - \lambda_\perp}{\lambda_\parallel + \lambda_\perp},$$

which, for a fixed value of $\lambda_\parallel/\lambda_\perp$, will be possible to satisfy over a range of $m$ centered on zero.

### 4.1 Constant-speed higher-order defects

The simplest manifestations of higher-order defects are those with constant rotational speed, which, in plane polar coordinates $(r,\theta)$ have the form $\psi = m\theta + \gamma$, and hence $\alpha(\theta,r) = \alpha_0(\theta) = (m-1)\theta + \gamma$. These patterns are also of particular interest because they have been experimentally realized.[29] In this case we have simply

$$\mathbf{b} + \mathbf{s} = \frac{m}{r}(\cos(2\alpha)\mathbf{e}_r + \sin(2\alpha)\mathbf{e}_\theta). \quad (21)$$

Away from the origin we have distributed Gauss curvature which arises, via the structural curvature, from the divergence of this term:

$$K_A = \frac{1}{2}(\lambda_\perp^{-2} - \lambda_\parallel^{-2})\nabla \cdot (\mathbf{b} + \mathbf{s})$$
$$= \frac{1}{2}(\lambda_\perp^{-2} - \lambda_\parallel^{-2})\frac{2(m-1)m}{r^2}\cos(2\alpha). \quad (22)$$

This distributed curvature vanishes for $m = 1$, as is familiar for log-spiral patterns. In contrast, all other constant-speed defects produce surfaces with distributed Gauss curvature $K_A \propto \cos(2\alpha) = \cos(2(m-1)\theta + 2\gamma)$, in agreement with the calculations in ref. 43 and 45.

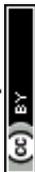

The integrated curvature at the defect point itself is characterized by eqn (20). Along an infinitesimal circle around the defect we have $\langle\cos(2\alpha_0)\rangle = 0$ for all $m \neq 1$, so the structural term vanishes leaving only the topological term. For $m = 1$ the angle $\alpha$ is constant, so $\langle\cos(2\alpha_0)\rangle = \cos(2\alpha)$ and both terms contribute. Within the circle we thus have

$$\int K_A \, dA_A = m\pi\left(1 - \frac{\lambda_\perp}{\lambda_\parallel}\right)\left(1 - \frac{\lambda_\parallel}{\lambda_\perp}\right) + \delta_{1m}\pi\left(\frac{\lambda_\parallel}{\lambda_\perp} - \frac{\lambda_\perp}{\lambda_\parallel}\right)\cos(2\alpha), \quad (23)$$

where the Kronecker $\delta_{1m}$ equals one when $m = 1$ and zero otherwise. Constant speed defects thus always encode intrinsically sharp points, with the sole exception of $m = 1$ defects at the critical angle $\alpha_c$ discussed previously.

This result highlights a profound difference between $m = 1$ and $m \neq 1$ constant-speed defects. For $m = 1$, the structural curvature gives zero distributed curvature, but produces a finite contribution to the integrated curvature at the point of the defect. For $m \neq 1$ the structural curvature encodes a pattern of distributed Gauss curvature that diverges in strength towards the defect point, but makes no contribution to the integrated curvature at the point itself because it cancels under integration due to its angular variation. A (heuristic) differential perspective on this difference is that taking the divergence in eqn (22) also produces a singular term at the origin, $(\lambda_\perp^{-2} - \lambda_\parallel^{-2})\pi m \cos(2\alpha)\delta^2(\mathbf{r})$. Then, integrating the combined (singular + distributed) divergence over an infinitesimal disk, the $\cos(2\alpha)$ variation causes the resulting structural curvature to vanish for all $m \neq 1$; but the singular term leads to a finite structural contribution for $m = 1$. This reflects a key symmetry difference in the patterns: adding an equal increment to $\alpha$ everywhere fundamentally changes the structure of an $m = 1$ pattern, but simply rotates an $m \neq 1$ pattern about the origin. Correspondingly, such an increment may change the integrated curvature at the defect point for $m = 1$, but not for $m \neq 1$.

### 4.2 Numerical calculation of activated surfaces for constant speed higher-order defects

Constant-speed higher-order defects thus offer a clear test of our main result, since the structural term accounts for the distributed curvature while the topological term accounts entirely for the finite integrated curvature at the defect point, and hence the intrinsically sharp point in the activated surface. We thus conduct numerical calculations using a bespoke shell-elasticity code (MorphoShell) to independently quantify their Gauss curvature and reveal their full activated shapes. Our approach closely follows the spirit of "non-Euclidian shells" with programmed metrics,[5] including both stretching and bending energies, with some specializations for incompressible rubber sheets.

In more detail, our elastic shell calculations describe the current shape of the sheet via the (3D) position $\mathbf{x}(\mathbf{r})$ of the material originally at the (2D) position $\mathbf{r}$ in the flat unactivated state. Using Latin indices for 3D and Greek for 2D, one can then compute the local deformation gradient $F_{i\alpha} = \partial_\alpha x_i$, and hence the local metric of the deformed sheet as $d\mathbf{x}^T d\mathbf{x} = d\mathbf{r}^T \mathbf{F}^T \cdot \mathbf{F} d\mathbf{r} \equiv d\mathbf{r}^T a d\mathbf{r}$.

If this current metric differs from the activated target metric $\bar{a} = (\lambda_\parallel^2 \mathbf{nn} + \lambda_\perp^2 \mathbf{n}^\star \mathbf{n}^\star)$ then the sheet must pay a stretching energy

$$E_S = \int \frac{\mu t}{2}\left(\text{tr}[a \cdot \bar{a}^{-1}] + \frac{1}{\det[a \cdot \bar{a}^{-1}]} - 3\right) dA, \quad (24)$$

which is simply the familiar stretching energy for a thin membrane of incompressible Neo-Hookean elastomer, with shear modulus $\mu$ and (unactivated) thickness $t$. This energy is minimized (and vanishes) if the sheet achieves an isometry, $a = \bar{a}$.

However, minimizing this stretching energy alone quickly reveals infinite numbers of non-smooth and unphysical isometries. To obtain sensible and physical surfaces, one must also include a sub-dominant bending energy, that penalizes extrinsic curvatures, encoded via the surface's second fundamental form, $B_{\alpha\beta} = (\partial_\alpha \partial_\beta \mathbf{x}) \cdot \hat{\mathbf{N}}$, where $\hat{\mathbf{N}}$ is the current surface normal. Following,[5] we use the standard bending energy for an incompressible sheet,

$$E_B = \int \frac{\mu t^3}{12 \det[\bar{a}]}\left(\text{tr}\left[\left(\bar{a}^{-1} \cdot B\right)^2\right] + \left(\text{tr}[\bar{a}^{-1} \cdot B]\right)^2\right) dA, \quad (25)$$

which is minimized when $B = 0$, (i.e., flat), as expected when the encoded metric does not vary through the thickness.

MorphoShell represents the deforming sheet via an unstructured triangulated mesh. The current metric for each triangle, $a$, requires first derivatives and is estimated from the unique linear deformation that describes the current positions of the triangle's three nodes: a standard constant-strain finite-element approach. The current second fundamental form, $B$, requires second derivatives and hence is estimated from the unique quadratic deformation consistent with the positions of six 'patch' nodes close to the triangle's centroid. Full details of these estimates are given in the ESI.† The activated form of the surface is then given by minimizing the total energy, $E_S + E_B$, over current node positions via either damped Newtonian dynamics or gradient descent.

The resultant surfaces for a range of 10 higher-order constant-speed defects are shown in Fig. 4. Since the defects form complicated 3D surfaces, additional viewpoints are provided in Fig. S1 and supplementary videos M1–M10 (ESI†). Our calculations are for free floating films, and have a large (rubber-like) actuation strains, leading to dramatic high-amplitude shapes. The results clearly exhibit the azimuthally oscillating distributed Gauss curvature given by eqn (22), leading to increasingly flowery surfaces at higher-orders, and good qualitative agreement with the experiments in ref. 29. As familiar from arrays of cones,[15] many of the defect shells can pop into different stable configurations. The configurations shown in Fig. 4 were selected as likely global energy minima, and some examples of alternatives are given in Fig. S2 (ESI†).

To verify our predicted structural and topological contributions to the Gauss curvature, we first compute the Gauss curvature of the activated surfaces, $K_A(r,\theta)$, via the angular deficit at each node in the mesh (see ESI† for details). In Fig. 5 we plot the local activated Gauss curvature around a ring of fixed large reference radius, $K_A(194t,\theta)$, for the $m = +5/2$ defect. Since this radius is far from the intrinsically sharp point at the origin, the local Gauss curvature stems only from the distributed contribution, and shows excellent agreement with the magnitude and form predicted in eqn (22).

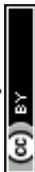

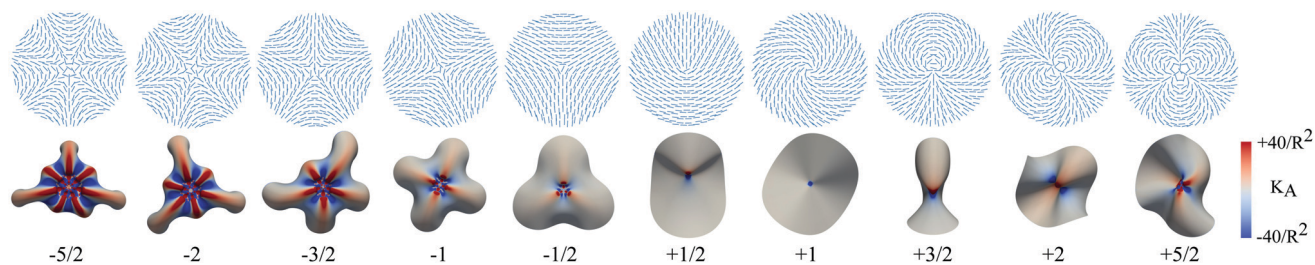

Fig. 4 Computational results for the activated morphologies of constant-speed defects in nematic elastomer sheets. The unactivated sheets are thin disks of radius $R$ and thickness $t \approx R/100$, as is typical for programmed nematic elastomer sheets. In all cases the programmed metric has $\lambda_{\parallel} = 0.75$, $\lambda_{\perp} = 1/\sqrt{\lambda_{\parallel}}$ except for $m = +5/2$, where $\lambda_{\parallel} = 0.9$ is depicted to avoid self-intersection. The $m = +1$ defect is programmed with $\alpha = \pi/4$, to isolate the topological contribution to the Gauss curvature, resulting in a weak anticone and enabling comparison with the higher-order defects.

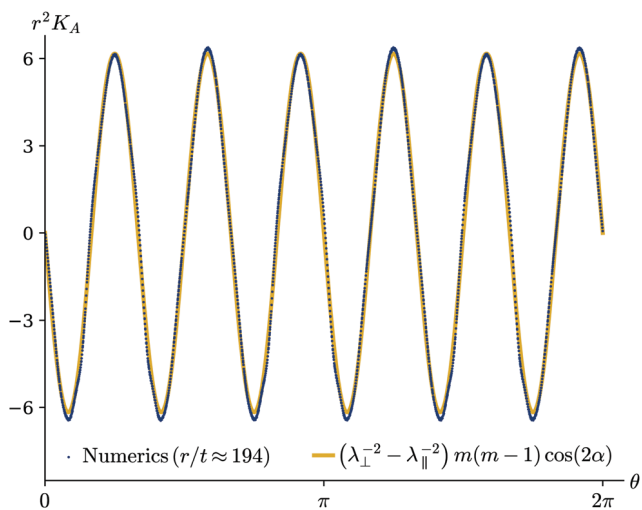

Fig. 5 Activated Gauss curvature as a function of $\theta$ at $r \approx 194t$ (far from the defect core), for an $m = -2$ constant-speed defect with $\gamma = \pi/4$ and thus $\cos(2\alpha) = \sin(6\theta)$. The values estimated numerically on a simulated surface show good agreement with eqn (22).

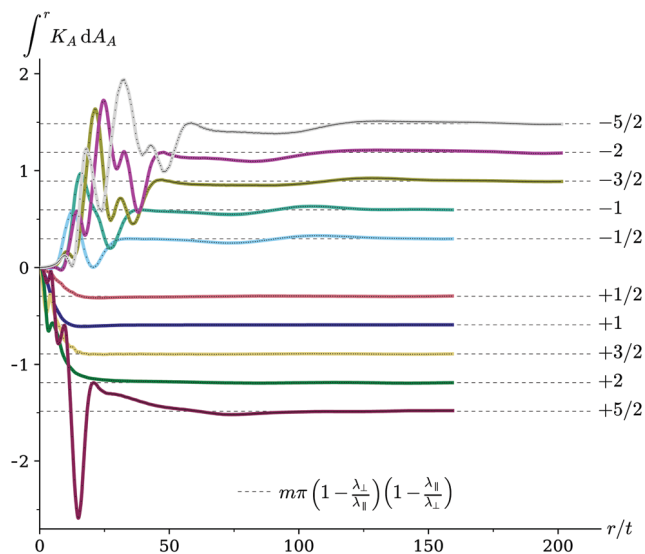

Fig. 6 Integrated Gauss curvature (up to radius $r$) for 10 simulated defects. Each defect generates the topological curvature predicted by eqn (23), forming a quantized ladder. Owing to finite-thickness effects, the curvature is not singularly concentrated at a point at the origin, but is spread over a finite core.

However, since the topological curvature is concentrated in a delta function at the origin, it cannot be interrogated by plotting $K_A$ directly. Rather, in the spirit of Gauss–Bonnet, in Fig. 6 we plot the integrated Gauss curvature within a reference-state disk centered on the origin, $\int_0^r \int_{\theta=0}^{2\pi} K_A \mathrm{d}A_A$, as a function of integration disk radius. In a perfect isometry the only contribution to the integrand would be a delta function of concentrated curvature at the center, leading the integral to have a constant value, $m\pi \left(1 - \frac{\lambda_{\perp}}{\lambda_{\parallel}}\right)\left(1 - \frac{\lambda_{\parallel}}{\lambda_{\perp}}\right)$, stemming entirely from the topological curvature. In practice, the bending energy prohibits a singular curvature at the origin, so this contribution is 'smeared out' over a finite non-isometric region around the origin. However, as one moves away from the origin, all the programmed curvature is indeed accounted for, and each integral asymptotes to the expected topological constant. We emphasize both the quantized nature of the ladder of asymptotes shown in Fig. 6, which reflects their topological origin, and also the different scalings with $\lambda_{\parallel}$ and $\lambda_{\perp}$ of the curvatures in Fig. 5 and 6, which highlight the distinction between the topological and structural contributions to the curvature.

The smeared-out regions near the origin in Fig. 6 involve a stretch-bend trade-off, resulting in locally non-isometric deformations that are beyond the scope of this paper. The extent of the smeared region is doubtless $\propto t$ (as explored for $m = +1$ cones in ref. 26), and would vanish in the vanishing thickness limit. In practice, the region extends a surprisingly large multiple of $t$ in higher-order defects, which we attribute to the very high distributed curvatures near the origin in these samples. However, ultimately Gauss–Bonnet guarantees that all the curvature encoded at the origin must be accounted for, once the boundary of the integration disk is large enough that the surface is isometric at the boundary.

## 5 Gauss curvature induced by directional fields with varying magnitudes

Finally, we consider a programmed directional growth field in which $\lambda_{\parallel}$ and $\lambda_{\perp}$ are also functions of position. Such programming has not yet been demonstrated in LCEs (although it could be achieved by additionally patterning the crosslink density or the strength of the order parameter), but is commonly found in biology: for example the exaggerated nectar spurs in Darwin's orchid develop *via* a process of cell elongation,[21] but the degree of elongation varies with distance from the tip of the spur. Given the local metric now has three degrees of freedom, $\psi$, $\lambda_\parallel$ and $\lambda_\perp$, it is possible to represent any programmed metric in this form. However, this is likely to be a natural representation in cases where the direction $\boldsymbol{n}$ is physically meaningful, such as being the direction of cell polarization. In this case we can again adopt the nematic coordinate system, $(u,v)$; the metric is still diagonal, but now $\lambda_\parallel$ and $\lambda_\perp$ are functions of $u$ and $v$. Applying eqn (2), the geodesic curvature is thus

$$\int k_{gA} dl_A = \int d\phi_A + \int \left( \frac{1}{\lambda_\parallel} \frac{\lambda_\perp \beta_u + \lambda_{\perp u}\beta}{\eta\beta} \beta dv - \frac{1}{\lambda_\perp} \frac{\lambda_\parallel \eta_v + \lambda_{\parallel v}\eta}{\eta\beta}\eta du \right). \quad (26)$$

To bring this into a coordinate-independent form, we again use $d\boldsymbol{l} = \eta \, du \, \boldsymbol{n} + \beta \, dv \, \boldsymbol{n}^*$ and recognize the (scalar) bend and splay, but must also recognize that $\frac{1}{\eta}\partial_u = \boldsymbol{n}\cdot\nabla$ and $\frac{1}{\beta}\partial_v = \boldsymbol{n}^*\cdot\nabla$, which allows us to eliminate the unknown fields $\eta$ and $\beta$ and write

$$\int k_{gA} dl_A = \int d\phi_A + \int \left( \frac{\lambda_\perp}{\lambda_\parallel} s \boldsymbol{n}^* + \frac{\boldsymbol{n}\cdot\nabla\lambda_\perp}{\lambda_\parallel}\boldsymbol{n}^* + \frac{\lambda_\parallel}{\lambda_\perp} b\boldsymbol{n} - \frac{\boldsymbol{n}^*\cdot\nabla\lambda_\parallel}{\lambda_\perp}\boldsymbol{n} \right) \cdot d\boldsymbol{l}. \quad (27)$$

As an aside, we note that this can be neatened up to give

$$\int k_{gA} dl_A = \int d\phi_A + \int \left( \frac{1}{\lambda_\parallel} \nabla\cdot(\boldsymbol{n}\lambda_\perp) \boldsymbol{n}^* - \frac{1}{\lambda_\perp} \nabla\cdot(\boldsymbol{n}^*\lambda_\parallel)\boldsymbol{n} \right) \cdot d\boldsymbol{l}. \quad (28)$$

However, returning to eqn (27) and again applying Gauss–Bonnet to a patch built from $u$ lines and $v$ lines, and containing defects with charge $m_i$, we can compute the contained Gauss curvature as

$$\int K_A dA_A = 2\pi \sum_i m_i + \oint \left( \frac{\lambda_\parallel}{\lambda_\perp} \boldsymbol{b} - \frac{\boldsymbol{n}^*\cdot\nabla\lambda_\parallel}{\lambda_\perp}\boldsymbol{n}^* - \frac{\lambda_\perp}{\lambda_\parallel}\boldsymbol{s} - \frac{\boldsymbol{n}\cdot\nabla\lambda_\perp}{\lambda_\parallel}\boldsymbol{n} \right) \cdot \hat{\boldsymbol{v}} dl. \quad (29)$$

If desired, eqn (12) can again be deployed to either bring the topological term inside the integral, or to evaluate the $\boldsymbol{b} - \boldsymbol{s}$ component of the integral. The resulting expressions are identical to eqn (13) and (14), except with an additional contribution, $-\int \left( \frac{\boldsymbol{n}^*\cdot\nabla\lambda_\parallel}{\lambda_\perp}\boldsymbol{n}^* + \frac{\boldsymbol{n}\cdot\nabla\lambda_\perp}{\lambda_\parallel}\boldsymbol{n} \right) \cdot \hat{\boldsymbol{v}} dl$, to the structural curvature arising from the varying magnitudes.

## 6 Discussion and conclusions

We have presented a simple result for the Gauss curvature developed by a flat sheet programmed with patterned directional growth. Our result is valid for all patterns of growth, even if they contain topological defects, and thus unifies previous (and superficially contradictory) results on continuously distributed Gauss curvature and singularly concentrated Gauss curvature at simple $+1$ defects, and extends these results to include all defect charges.

As seen most clearly in eqn (13), our result reveals a subtle interplay between the topology of defects and the precise structure of a given defect's realization, which in turn generates a subtle interplay between topological defects and intrinsically sharp points. The topological term in eqn (13) contributes a finite integrated curvature to the defect point, which depends only on the charge of the defect and the value of $\lambda_\parallel/\lambda_\perp$. In particular, the topological term always gives negative contributions for positive charge defects and *vice versa*, and is invariant under taking the orthogonal dual of a pattern. In contrast, the structural term in eqn (13) can produce both distributed Gauss curvature, and contribute finite integrated curvature to points with discontinuous director. At topological defects, the director is discontinuous, and the structural term makes a finite contribution to the defect point. This structural contribution depends on the exact spatial form of the defect pattern, and is always inverted, $K_{str} \to -K_{str}$, by taking the orthogonal dual of a pattern. As seen in eqn (20), the resultant integrated curvature at a point with a topological defect contains both structural and topological contributions, and can thus lie anywhere in a given range, centered on the topological term and spanned by the structural term. Generically the two contributions do not exactly cancel, so the defect point has a finite integrated curvature and encodes an intrinsically sharp point in the activated surface. However, it is possible to structure defects such that the two contributions exactly cancel. In this case the defect point carries zero integrated curvature and does not create an intrinsically sharp point in the activated surface.

Conversely, although all the intrinsically sharp points we have covered are generated at topological defects, it appears possible to program one *via* the structural curvature alone, at a point without a topological defect. A full treatment of the conditions required for such points lies beyond the scope of this paper. However, heuristically, the structural curvature within a patch is controlled by $\oint (\boldsymbol{b}+\boldsymbol{s})\cdot\hat{\boldsymbol{v}} dl$, which may encode a finite value at any point with divergent $\boldsymbol{b}$ or $\boldsymbol{s}$. Since bend and splay are the curvatures of director lines and orthogonal dual lines respectively, such a divergence certainly requires infinitely curved director/dual lines, and hence a discontinuity in director at the point. However, such a discontinuity would arise at any kink in a director curve, which need not carry a topological charge. An intrinsically sharp point thus requires a point of discontinuous director, but the discontinuity need not be a topological defect.

The key enabling approach to characterizing the curvature encoded at defects has been to use Gauss–Bonnet to evaluate the Gauss curvature within a patch in terms of its boundary properties, resulting in a formula for the Gauss curvature that needs integrating along the boundary but not over the (defect-containing) area of the patch itself. In the particularly simple case where the patch boundary is constructed from $u$ and $v$ lines (integral curves of $\boldsymbol{n}$ and $\boldsymbol{n}^*$) we can go a step further and conduct the boundary integral along each segment to calculate the Gauss curvature in terms of properties at the corners. More precisely, we can conduct the integral in eqn (14) along

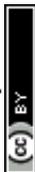

each segment, which yields the change in angle of the path segment, so the Gauss curvature within is

$$\int K_A dA_A = \left(1 - \frac{\lambda_\parallel}{\lambda_\perp}\right)\sum \Delta\psi_u + \left(1 - \frac{\lambda_\perp}{\lambda_\parallel}\right)\sum \Delta\psi_v,$$

where $\Delta\psi_u$ and $\Delta\psi_v$ are the change in director angle (and hence change in path angle) along the $u$ line and $v$ line segments. The Gauss curvature within the patch is thus entirely determined from the director angles at the corners. For example, for the quadrangular path shown in Fig. 2, the Gauss curvature within is simply

$$\int K_A dA_A = \left(\frac{\lambda_\perp}{\lambda_\parallel} - \frac{\lambda_\parallel}{\lambda_\perp}\right)(\psi_a - \psi_b + \psi_c - \psi_d).$$

Going further, if the boundary is largely composed of $u$ lines, the Gauss curvature within will be $\int K_A dA_A \to \left(1 - \frac{\lambda_\parallel}{\lambda_\perp}\right)2\pi m$, while if it is largely composed of just $v$ lines it will be $\left(1 - \frac{\lambda_\perp}{\lambda_\parallel}\right)2\pi m$: results that are familiar from radial and circumferential +1 defects, but now generalized to patches containing multiple higher-order defects with the correct shape of boundary.

We anticipate that these geometric insights will be of use for designing patterns of growth to achieve particular surfaces. Current strategies for inverse design cannot produce patterns containing defects,[40] but our results highlight that defects not only offer a route to designing surfaces with sharp features, but can also occur without any concentrated Gauss curvature as part of the programming of a smoothly curved surface. Furthermore, our results show that the integrated Gauss curvature of a surface is intimately related to the total topological charge it contains, suggesting that the inclusion of topological defects may be crucial to the design of surfaces containing large integrated amounts of Gauss curvature, such as needed to wrap (or even double-wrap) a sphere.

Looking beyond our current results, we first note the possibility of encoding growth into an initially curved surface, so that it morphs into a different curved surface on activation. This situation is actually typical in biology, and, via "4D" printing, is increasingly accessible to engineers.[30] Although we defer a full investigation to a later date, we anticipate that our existing approach will generalize straightforwardly; indeed, as long as one uses nematic coordinates, the metric of an initially curved surface could be represented via eqn (3) without modification suggesting that many of our current results will be directly applicable. A second extension would be to compute Gauss curvature when the director is discontinuous along a seam, rather than a simple point defect.[45,49,50] Our current approach relies on the director being continuous on a patch boundary, then using Gauss–Bonnet to capture the curvature within. Our results thus apply directly to a seam that lies entirely within a patch (rather like a Kirigami cut) which can be circumnavigated without difficulty, and an appropriate $m$ assigned from the winding of $\boldsymbol{n}$ around the boundary. However, if the seam passes through the patch boundary then director discontinuity also occurs on the boundary, and the concept of enclosed topological charge breaks down. As we will show in a forthcoming paper, in this case the Gauss–Bonnet approach still works, but yields rather different results.

Finally, we comment that many biological tissues also have nematic or polar order (stemming from the anisotropy of their cells), and there is growing appreciation of the physiological implications of defects: for example +1/2 and −1/2 defects in epithelial tissue are known to be drive cell shedding.[51] Similarly the patterns of directional elongation that generate both the exaggerated nectar spurs in Darwin's orchid[21] and the reproductive whorl in the green alga Acetabularia acetabulum[22,25] are centered on +1 defects, and these must be accounted for properly to compute the Gauss curvature and understand the shape programming. Unlike in LCEs, biological growth fields tend to be patterned in both direction and magnitude, but fortunately our results straightforwardly generalize to this case (eqn (29)). Furthermore, given the metric itself has quadrupolar symmetry, this formulation is not limited to tissues with nematic order, but encompasses the full spectrum of encodable metrics, and is likely to be useful in polar tissues with topological defects as well as their nematic counterparts.

## Conflicts of interest

There are no conflicts of interest to declare.

## Acknowledgements

J. S. B. is supported by a UKRI Future Leader Fellowship, grant no. MR/S017186/1. D. D. is supported by the EPSRC Centre for Doctoral Training in Computational Methods for Materials Science, grant no. EP/L015552/1, and thanks the creators of Distmesh.[52]

## Notes and references

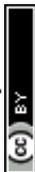

# Supplementary Information
for 'Defective nematogenesis: Gauss curvature in programmable shape-responsive sheets with topological defects'

Daniel Duffy (dld34@cam.ac.uk)

Dr John S. Biggins

## S1  Supplementary figures

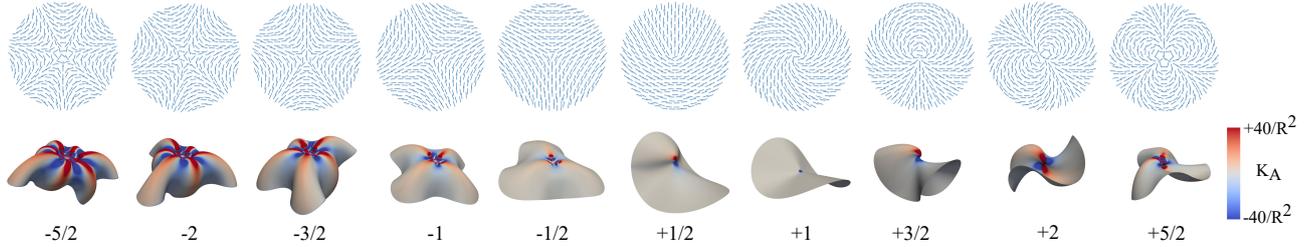

Figure S1: Alternative views of the simulated surfaces in Fig. 4 of the main paper. Note that despite the $2(1-m)$-fold rotational symmetry of the $m < 0$ patterns, this symmetry is broken by the observed shapes, even though all the large 'lobes' popped the same way. The same is true for $m = +5/2$, and is expected to hold for greater $m$ values too. See also the supplementary movies M1-10.

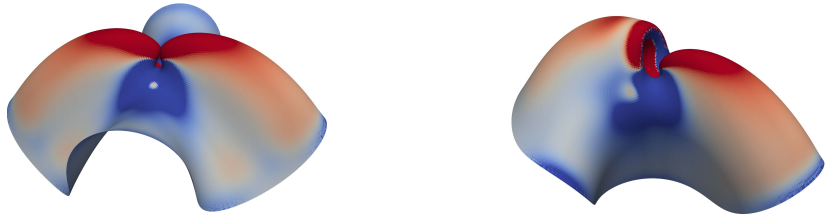

Figure S2: Two other simulations of the $+2$ defect, showing that a variety of different shapes are possible if the sheet is allowed to 'pop' in different ways. These shapes had a higher residual energy than the $+2$ defect in Fig. S1, suggesting that they are local energy minima, while that in Fig. S1 is expected to be the global minimum. These might nonetheless be realised physically however.

## S2  Computational details

Computations were carried out on disks of (reference-state) thickness $t$ and radius $R \gg t$, programmed with constant-speed defect metrics with $\lambda_\perp = 1/\sqrt{\lambda_\parallel}$. The disks were meshed with 2D triangles of edge length $h \lesssim t$ to resolve the largest curvatures. All meshes were created with the PyDistMesh implementation of the DistMesh algorithm [1]. The specific parameters for each disk are given in the table below; the finest mesh had $> 1.8$ million triangles.

| Simulation(s) | $\lambda_\parallel$ | $h/t$ | $R/t$ |
|---|---|---|---|
| Fig. 4, 6: $-5/2$, $-3/2$ | 0.75 | 0.43 | 217 |
| Fig. 4, 6: $-2$ | 0.75 | 0.87 | 217 |
| Fig. 4, 6: $-1 \to +2$ | 0.75 | 0.87 | 173 |
| Fig. 4: $+5/2$ | 0.9 | 0.95 | 190 |
| Fig. 5: $-2$ | 0.75 | 0.87 | 260 |
| Fig. 6: $+5/2$ | 0.75 | 0.43 | 173 |

The initial states were perturbed slightly in the out-of-plane direction to encourage particular 'popping' patterns that are likely global energy minima. The shells are multistable and various other stable, differently-popped solutions were also encountered (see Fig. S2 above).

MorphoShell calculates the metric and second fundamental form for each triangle based on current node positions (see Section S3). The force on each node is then calculated as the gradient of the total energy ((24)+(25) in the main paper), and the nodes are moved accordingly via damped Newtonian dynamics or gradient descent. In our calculations equilibrium was deemed achieved when non-dimensionalized elastic force and speed both fell below $5 \times 10^{-5}$. The MorphoShell code and associated files used to create Fig. 6 (main paper) are available as supplementary material. See DOI: `https://doi.org/10.17863/CAM.58084`

Radius $R$ for each disk was chosen to be large enough that the integrated Gauss curvature attained its asymptotic far-field value away from the core (Fig. 6, main paper). Non-Euclidean shells are also known to have a non-isometric boundary layer within $\sim t$ of the outer perimeter, which would obscure the asymptotic behaviour, so the plots in Fig. 6 (main paper) are truncated before the boundary layer is reached. Furthermore the sheet was made thinner (by a factor $\sim 15$) and stiffer within $\sim 10t$ of the boundary (without altering the bending modulus), to restrict the boundary layer to this region.

## S2.1 Gauss curvature estimates for Figs 5, 6 of main paper

Figs. 5, 6 of the main paper required a robust estimate of Gauss curvature, on a triangulated mesh. We borrowed the well-known 'angle deficit' method from discrete differential geometry. This method is motivated by an exact discrete analogue of the Gauss-Bonnet theorem for a triangulated mesh surface $S$ [2]:

$$\sum_{i \in interior} (2\pi - \Xi_i) + \sum_{i \in boundary} (\pi - \Xi_i) = 2\pi \chi(S), \tag{1}$$

where the sums are over the non-boundary and boundary nodes respectively, and

$$\Xi_i = \sum_\Delta \beta_\Delta \qquad (2)$$

is the sum of mesh-triangle interior angles $\{\beta_\Delta\}$ around the node $i$. Comparing with the continuum Gauss-Bonnet theorem, this provides a natural definition for the integrated Gauss curvature of an interior node in a triangulated mesh: the 'angle deficit', $2\pi - \Xi_i$. This is a widely used Gauss curvature measure [3], and is the measure used in Fig. 6 (main paper).

For Fig. 5 (main paper), the local Gauss curvature was also required. To obtain this, we simply divide each node's angle deficit by an area assigned to that node, taken to be $A_i = \sum_\Delta A_\Delta/3$, so that each triangle contributes $1/3$ of its area to each vertex. Thus we obtain:

$$K_i = \frac{2\pi - \Xi_i}{A_i}. \qquad (3)$$

This and similar estimates for $K$ are not locally convergent in full generality [4]. However they are unproblematic for highly regular meshes such as ours, (hence their widespread use), and the estimate $K_i$ performed well in our case.

Fig. 5 (main paper) is a scatter plot of the above estimate $K_i$, for all nodes in a reference-state annulus with centre-line radius $r \approx 194\,t$ and extent $\Delta r \approx 1.3\,t$.

## S3  Numerical estimates of the metric and second fundamental form

MorphoShell evolves an unstructured triangulated mesh. Each node has a fixed 2D position vector $^{(i)}\boldsymbol{r}$ in the initial (reference) state, and a dynamical 3D position vector $^{(i)}\boldsymbol{x}$ in the deformed ('current') state.

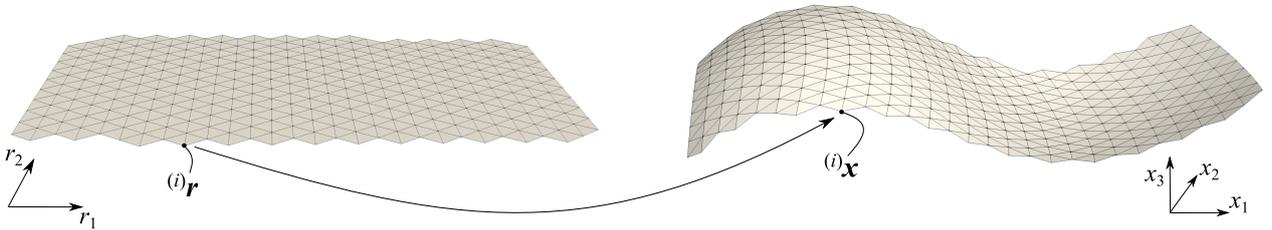

Figure S3: The reference and deformed states of a small piece of mesh.

The deformed metric for each triangle in the mesh is taken to be the unique 2x2 matrix mapping its initial squared side lengths to their deformed values.

We use a 'patch fitting' approach to estimate the second fundamental form, as is common in computer graphics. In contrast to some others [5], our estimate contains only node position degrees of freedom.

For a given triangle $T$, six 'patch nodes' are determined: $T$'s three vertices, and three further nodes selected from nearby triangles. The deformed surface in the vicinity of the triangle is approximated by a 'patch' surface $\boldsymbol{p}(\boldsymbol{r})$ that is a quadratic function of Cartesian reference coordinates $(r_1, r_2)$:

$$\boldsymbol{p} = \begin{pmatrix} A_1 & B_1 & C_1 & D_1 & E_1 & F_1 \\ A_2 & B_2 & C_2 & D_2 & E_2 & F_2 \\ A_3 & B_3 & C_3 & D_3 & E_3 & F_3 \end{pmatrix} \begin{pmatrix} 1 \\ \Delta r_1 \\ \Delta r_2 \\ (\Delta r_1)^2 \\ \Delta r_1 \Delta r_2 \\ (\Delta r_2)^2 \end{pmatrix} \equiv \boldsymbol{\Lambda} \begin{pmatrix} 1 \\ \Delta r_1 \\ \Delta r_2 \\ (\Delta r_1)^2 \\ \Delta r_1 \Delta r_2 \\ (\Delta r_2)^2 \end{pmatrix}, \tag{4}$$

where $\Delta r_1 \equiv r_1 - \mathring{r}_1$ and $(\mathring{r}_1, \mathring{r}_2)$ are the reference coordinates of $T$'s centroid. This is nothing more than a truncated Taylor expansion for each component of $\boldsymbol{p}(\boldsymbol{r})$, valid as long as the surface's local radius of curvature is significantly larger than the element size.

Requiring the patch surface to pass through the current coordinates of the six patch nodes then provides six constraints that can be used to find $A_1, ..., F_3$, the elements of $\boldsymbol{\Lambda}$:

$$\boldsymbol{M\Lambda}^T \equiv \begin{pmatrix} 1 & {}^{(1)}\Delta r_1 & {}^{(1)}\Delta r_2 & ({}^{(1)}\Delta r_1)^2 & ({}^{(1)}\Delta r_1)({}^{(1)}\Delta r_2) & ({}^{(1)}\Delta r_2)^2 \\ 1 & {}^{(2)}\Delta r_1 & {}^{(2)}\Delta r_2 & ({}^{(2)}\Delta r_1)^2 & ({}^{(2)}\Delta r_1)({}^{(2)}\Delta r_2) & ({}^{(2)}\Delta r_2)^2 \\ 1 & {}^{(3)}\Delta r_1 & {}^{(3)}\Delta r_2 & ({}^{(3)}\Delta r_1)^2 & ({}^{(3)}\Delta r_1)({}^{(3)}\Delta r_2) & ({}^{(3)}\Delta r_2)^2 \\ 1 & {}^{(4)}\Delta r_1 & {}^{(4)}\Delta r_2 & ({}^{(4)}\Delta r_1)^2 & ({}^{(4)}\Delta r_1)({}^{(4)}\Delta r_2) & ({}^{(4)}\Delta r_2)^2 \\ 1 & {}^{(5)}\Delta r_1 & {}^{(5)}\Delta r_2 & ({}^{(5)}\Delta r_1)^2 & ({}^{(5)}\Delta r_1)({}^{(5)}\Delta r_2) & ({}^{(5)}\Delta r_2)^2 \\ 1 & {}^{(6)}\Delta r_1 & {}^{(6)}\Delta r_2 & ({}^{(6)}\Delta r_1)^2 & ({}^{(6)}\Delta r_1)({}^{(6)}\Delta r_2) & ({}^{(6)}\Delta r_2)^2 \end{pmatrix} \boldsymbol{\Lambda}^T \tag{5}$$

$$= \begin{pmatrix} {}^{(1)}x_1 & {}^{(1)}x_2 & {}^{(1)}x_3 \\ {}^{(2)}x_1 & {}^{(2)}x_2 & {}^{(2)}x_3 \\ {}^{(3)}x_1 & {}^{(3)}x_2 & {}^{(3)}x_3 \\ {}^{(4)}x_1 & {}^{(4)}x_2 & {}^{(4)}x_3 \\ {}^{(5)}x_1 & {}^{(5)}x_2 & {}^{(5)}x_3 \\ {}^{(6)}x_1 & {}^{(6)}x_2 & {}^{(6)}x_3 \end{pmatrix} \equiv \boldsymbol{D}, \tag{6}$$

where for example ${}^{(5)}x_2$ is the second deformed-state coordinate of the fifth patch node. Thus

$$\boldsymbol{\Lambda} = \boldsymbol{D}^T \boldsymbol{M}^{-T}. \tag{7}$$

Note that $\boldsymbol{M}^{-T}$ is a reference-state quantity, so it can be computed just once and re-used, while $\boldsymbol{D}^T$ is updated with the new node positions at each step.

Using $B_{\alpha\beta} = (\partial_\alpha \partial_\beta \boldsymbol{x}) \cdot \hat{\boldsymbol{N}}$, we combine $\boldsymbol{\Lambda}$ with the triangle face normal $\hat{\boldsymbol{N}}$ to give an estimate

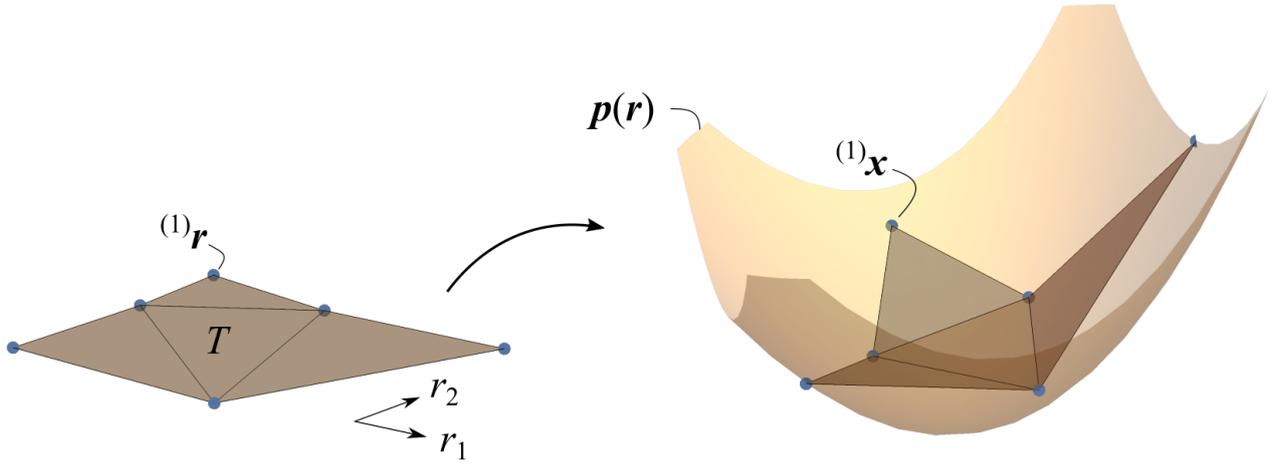

Figure S4: Sketch of the reference and deformed state for a triangle $T$ and its six patch nodes, and the fitted patch surface $\boldsymbol{p}(\boldsymbol{r})$.

of the second fundamental form for triangle $T$:

$$\boldsymbol{B} = \sum_{j=1}^{3} \begin{pmatrix} 2\,\Lambda_{j4}\,\hat{N}_j & \Lambda_{j6}\,\hat{N}_j \\ \Lambda_{j6}\,\hat{N}_j & 2\,\Lambda_{j5}\,\hat{N}_j \end{pmatrix}. \qquad (8)$$

The three non-vertex patch nodes were simply chosen to be those closest to $T$'s centroid in the reference state, subject to $\boldsymbol{M}^{-T}$ being suitably well conditioned.